\documentclass{article}
\usepackage{pstricks,psfrag}
\usepackage{xcolor,rotating,epic,eepic}
\usepackage{amssymb}
\usepackage{amsmath}
\usepackage{times} 
\usepackage{epsfig,t1enc}
\usepackage{graphicx}
\usepackage{graphics}
\newtheorem{definition}{Definition}
\newtheorem{theorem}{Theorem}
\newtheorem{lemma}{Lemma}

\def\II{\hbox{1\kern-.2em\hbox{I}}}

\def\Del{Del}
\def\Gab{Gab}
\def\Rng{Rn}

\title{Continuum Percolation in the Relative Neighborhood Graph} 
\author{Billiot, J.-M. \and Corset, F. \and Fontenas, E.\\
{\begin{small}LJK, SAGAG Team, BSHM, Universit\'e Pierre Mend\`es France\end{small}}\\
{\begin{small}1251 avenue centrale, B.P. 47, 38040 Grenoble cedex 9, France\end{small}}\\
{\begin{small}jean-michel.billiot@upmf-grenoble.fr\end{small}}
}

\begin{document}
\maketitle
\begin{abstract}
In the present study, we establish the existence of nontrivial site percolation threshold in the Relative Neighborhood 
Graph (RNG) for Poisson stationary point process with unit intensity in the plane.  
\end{abstract}



\section{Introduction}

Percolation theory is very useful to describe various physical
phenomena. In particular, there are 
important connections with phase transition problems~\cite{Grimmet99,Georgii00}.

The interest for percolation problems has
grown rapidly during the last decades: see Lyons and Peres~\cite{Lyons02} for percolation on trees and networks  
and Meester and Roy~\cite{Meester96} for continuum percolation and the references therein. 
In 1996, H{\"a}ggstr{\"o}m and Meester \cite{Haggstrom96} proposed results for
continuum percolation problems for the $k$-nearest neighbor graph under Poisson process. 
In a recent paper, Balister and Bollob\'as \cite{Balister10} give bounds on $k$ for the $k$-nearest neighbor graph 
for percolation with several possible definitions.

Site and bond percolation as well as phase transition for several statistical mechanics models have also been studied by H{\"a}ggstr{\"o}m \cite{Haggstrom00} for general graphs and in particular for the Delaunay triangulation for the Poisson point process. It also shows for graphs with a bounded degree if the bond percolation threshold is not trivial then it implies a phase transition for the Ising model on this graph.  
For general results on Delaunay graphs and Voronoi tesselations see Moller~\cite{Moller94}. 
For recent results on percolation on these graphs see Balister et al.~\cite{Balister5,Balister09}.\\

Benjamini and Schramm \cite{Benjamini96} proposed a comprehensive study on general graphs, with special focus on Cayley graphs, quasi-transitive graphs and planar graphs.
Recently, Procacci and Scoppola \cite{Procacci04} proposed sufficient conditions on infinite graphs to deduce a non trivial bond percolation threshold. Among these assumptions, they assume that the dual graph is bounded degree. 
It is interesting to relax this condition in order to deal with proximity random
graphs (which in general have a dual not bounded degree) like the skeletons on point
Poisson processes of the plane. These random graphs includes the Gabriel graph and the Relative Neighborhood Graph (RNG) which are important for many applications.


Remark that continuous models defined on nearest neighbors graphs are interesting for small temperature as alternative of standard models on regular networks, because it allows vibrations and deformations of the network and may be find an application in physics of the solid state. In particular one example is the study of the order-disorder transition of binary alloys or ionic cristals. It is well-known that Delaunay graph or Voronoi regions
 (rather called Wigner-Seitz grid and Brillouin zone in physics framework) take a fondamental place for the understanding of the electrical current, waves propagation and phase transitions observed by Bragg diffraction of X rays.

Another domain of application
should be found in cancerology for the study of the growth of tumour when the cancer cells
suddently begin to invade healthy tissue. The Delaunay graph is well adapted for such
study as explained in~\cite{BBD1} but the RNG should give some more information.
More precisely, in histology
we have some slides with marked cells (cancer cells and normal cells in first approximation)
and probably, a connection between percolation and the grade
of a given cancer will be helpful to give an aid for the diagnostic of a pathologist.

The existence 
and unicity at small activity of nearest neighbors stationary Gibbs states can be found in Bertin et al.~\cite{BBD4,BBD3,BBD2}.
In Bertin et al.~\cite{BBD5}, the phase transition in the Delaunay continuum Potts model is established.
It is a generalization of the Lebowitz-Lieb  model as described in Georgii and H\"aggstr\"om~\cite{Georgii96}
where the soft repulsion between several species of particules acts on the Delaunay graph.
What is the good Delaunay subgraph on which the repulsion is strong enough
to maintain a phase transition? In terms of percolation, it means: is bond 
percolation maintained in this subgraph? 
Bertin et al.~\cite{BBD6} gave an answer for the Gabriel graph. Another well-known subgraph of Delaunay graph is the Minimum Spanning Tree (MST).
The structure of spanning forest generalized minimum spanning tree for infinite graphs.
Constructed with the greedy algorithm, this graph exactly is described in the book of Meester and Roy~\cite{Meester96}. The link between branching number and percolation on trees is proved by Lyons~\cite{Lyons90}.
Connection between minimum spanning forest and occupied and vacant percolation
is strong (for more details on simultaneous uniqueness see Alexander~\cite{Alexander95}). 
Thus, the minimum spanning forest is a tree with one infinite path a.s. 
in two dimensions when the points are distributed under a stationary Poisson process.
It comes that a.s. the site or bond critical threshold are equal to $1$.

The present study gives an answer for the RNG, well known in computational geometry, when the points are distributed under a stationary Poisson point process with unit intensity in the plane. 
We adapted a powerful method of the rolling ball proposed by Balister and Bollob\'as~\cite{Balister10} relying on 1-independent bond percolation on $\mathbb{Z}^2$. 
Then if we control the probability of having less than a fixed number of points in a given region and considering the event that all the sites are open in this region, 
we can proceed similarly as in H\"aggstr\"om~\cite{Haggstrom00}.\\

The paper is organized as follows. The first section is devoted to some definitions and notations. 
Next, the main result on site and bond percolation on the RNG is given. 
Then, we give the proof of the main result, by using a result of bond percolation in the 1-independent case in $\mathbb{Z}^2$ and the rolling ball method. 
We conclude on possible extensions of this work.

\section{Notations and definitions}
 
Let $|A|$ denotes the Lebesgue measure
if the set $A$ is a bounded Borel set of $\mathbb{R}^2$, and
the counting measure if $A$ is a discrete set.

Given a finite box $\Lambda\subset\mathbb{R}^2$, 
we denote by $\Pi_{\Lambda}$ the Poisson point process on the locally finite 
set of points in $\Lambda$ denoted by ${\Omega}_\Lambda$ with intensity $1$ i.e.
$$\int_{{\Omega}_\Lambda} f d\Pi_{\Lambda}=\exp\left(-|\Lambda|\right)\sum\limits_{n=0 }^\infty\frac{1}{n!} \int_{\Lambda^n} f(\{x_1,\ldots x_n\})dx_1,\ldots dx_n$$
for any bounded measurable function f on $\Omega_\Lambda$.

Let $\Omega$ the set of locally finite subsets of $\mathbb{R}^2$.
We have to consider only the configurations $\varphi \in \Omega$ which are in general position
(four points on the same circle do not occur and no three points are colinear) in order to ensure the existence and unicity 
of the Delaunay graph. One can notice that, for any stationary point processes,
the probability of the set of tesselations in general position is equal to one~\cite{Moller94}. Let us recall the definitions of Delaunay, Gabriel and relative neighborhood graphs.

\begin{definition}
\label{delaunay} The Delaunay graph $\Del_2(\varphi)$ of a configuration $\varphi$ in $\Omega$
is the set of edges of the unique triangulation $\Del_3(\varphi)$ in which the interior of the circle
circumscribed by every triangle of $\Del_3(\varphi)$ does not contain any point of $\varphi$.
\end{definition}

\begin{figure}[ht!]
\centering
\includegraphics[scale=0.5]{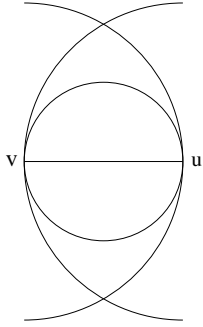}
\caption{Vacuity regions for Gabriel and Relative neighborhood graphs for an edge $uv$}
\label{lune}
\end{figure}

\begin{definition}
\label{gabriel} 
The Gabriel graph $\Gab(\varphi)$ of a configuration $\varphi$ in $\Omega$
is defined as the set of edges $\{u,v\}\subset\varphi$ such that the open circle with $\{u,v\}$ as diameter 
does not contain any point of the configuration $\varphi$.
\end{definition}
\begin{definition}
\label{Rn}
The Relative Neighborhood graph $\Rng(\varphi)$ of a configuration $\varphi$ in $\Omega$
is defined as the set of edges $\{u,v\}\subset\varphi$ such that the intersection of the disks with center $u$ and $v$ with radius $uv$  does not contain any point of the configuration $\varphi$.
\end{definition}
These graphs are planar in $\mathbb{R}^2$. Furthermore, the RNG is a 
subgraph of the Gabriel graph which is a subgraph of the Delaunay Graph (see figure~\ref{lune}).

\section{Main result}
\label{sec-PGG}

We first deal with the site percolation on the RNG for the Poisson point process $\Phi$. 
We introduce the Bernoulli process $\Theta(\Phi)$ providing the type picking mechanism ($1$ 
for open and $0$ for closed) of the points (or sites) in $\Phi$. Obviously, $\mathbf{\Phi}=(\Phi,\Theta(\Phi))$ 
can be seen as a marked Poisson process. The probability measure of $\mathbf{\Phi}$ is given by~:
$$\mathbb{P}(d\varphi,d\theta)=\int \Pi(d\varphi)\mu_\varphi^{p}(d\theta)$$
where $\mu_\varphi^{p}$ is the probability measure on $\{0,1\}^{\varphi}$ of the 
Bernoulli process $\Theta(\varphi)$ given the configuration $\varphi \in \Omega$. Similarly $\mu_{\Rng(\varphi)}^{p}$ is the probability measure on $\{0,1\}^{\Rng(\varphi)}$ of the 
Bernoulli process $\Theta(\Rng(\varphi))$ given the graph $\Rng(\varphi)$ where $\varphi\in\Omega$. We define $p_{c}^{site}(\varphi,\Rng(\varphi))$ and $p_{c}^{bond}(\varphi,\Rng(\varphi))$ such that:
$$\mu_\varphi^{p}(\exists\,\,\mbox{at least one infinite open cluster in }\,\Rng(\varphi) )$$
$$=\left\{\begin{array}{ll}1&{\rm if}\quad p>p_{c}^{site}(\varphi,\Rng(\varphi))\\
0&{\rm if}\quad p<p_{c}^{site}(\varphi,\Rng(\varphi))\end{array}\right.$$
$$\mu_{\Rng(\varphi)}^{p}(\exists\,\,\mbox{at least one infinite open cluster in }\,\Rng(\varphi) )$$
$$=\left\{\begin{array}{ll}1&{\rm if}\quad p>p_{c}^{bond}(\varphi,\Rng(\varphi))\\
0&{\rm if}\quad p<p_{c}^{bond}(\varphi,\Rng(\varphi))\end{array}\right.$$
Let us recall \cite{Grimmet99} the following well known relation between $p_{c}^{site}$ and $p_{c}^{bond}$ on a given graph with bounded degree: as the graph $(\varphi,\Rng(\varphi))$ have a degree at most $6$, $$\forall\varphi\in \Omega,\quad
1/5\leq p_{c}^{bond}(\varphi,\Rng(\varphi))\leq p_{c}^{site}(\varphi,\Rng(\varphi))\leq 1-[1-p_{c}^{bond}(\varphi,\Rng(\varphi))]^{6}.$$

We now introduce $p_c^{\mathrm{site}}\left(\Rng,\Pi\right)$ defined as the lowest $p$ for which the probability of the event that there exists an infinite open cluster in the RNG relative to the marked point Poisson process $\mathbf{\Phi}$ is equal to $1$. By ergodicity
of $\Pi$, this previous event, invariant by translation, is a trivial event. The marked point Poisson process $\mathbf{\Phi}$ may exhibit some percolation phenomenon with critical value $p_c^{\mathrm{site}}\left(\Rng,\Pi\right)$. We want to prove in the following theorem that this is a non trivial critical value, i.e., 

\begin{theorem} \label{prop-pg} $p_c^{\mathrm{site}}\left(\Rng,\Pi\right)<1 \text{ and }p_c^{\mathrm{bond}}\left(\Rng,\Pi\right)<1.$
\end{theorem}

The following section is devoted to the proof of this theorem.

\section{Proof}
\begin{figure}[ht!]
\centering
\includegraphics[scale=0.5]{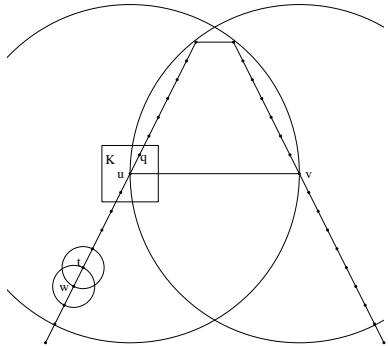}
\caption{Example of construction of configuration with arbitrary large \Rng- path.}
\label{rngh}
\end{figure}

First, we point out that the method proposed by H\"aggstr\"om~\cite{Haggstrom00} and used for $k$ nearest neighbor graph and Delaunay graph, 
as well as the adaptation by Bertin et al.~\cite{BBD6} to the Gabriel graph does not applied to the RNG. However, it is not sufficient 
to control the probability of having at least one point and less than a fixed number of points in each small box $K$ (see figure \ref{rngh}). 

Indeed, we may choose some configuration of points such that the length and number of points of a $\Rng-$ path between two points are 
arbitrary large, see figure \ref{rngh}. 
$wt$ is an edge of the RNG but $uv$ is not because for example the point $q$ belongs to the vacuity region of this edge.

\begin{figure}
\centering
\unitlength 1cm
\begin{picture}(11,6)(0,2)%
\color{black}
\thinlines 

\path(1,7)(1,3)(5,3)(5,7)(9,7)(9,3)
\path(1,7)(5,7)
\path(5,3)(9,3)
\put(3,5){\circle{2}}
\put(7,5){\circle{2}}
\path(3,6)(7,6)
\path(3,4)(7,4)
\dottedline[\circle*{.01}]{0.1}(6.25,5)(6.2418,5.1279)(6.2173,5.2537)(6.1769,5.3753)(6.1213,5.4907)(6.0514,5.5981)
(5.9683,5.6957)(5.8735,5.7818)(5.7684,5.8551)(5.6548,5.9144)(5.5345,5.9587)(5.4096,5.9872)(5.2821,5.9995)
(5.154,5.9954)(5.0275,5.9749)(4.9046,5.9385)(4.7875,5.8866)(4.6779,5.8202)(4.5777,5.7403)(4.4886,5.6482)
(4.4119,5.5455)(4.349,5.4339)(4.3009,5.3151)(4.2684,5.1912)(4.2521,5.0641)(4.2521,4.9359)(4.2684,4.8088)
(4.3009,4.6849)(4.349,4.5661)(4.4119,4.4545)(4.4886,4.3518)(4.5777,4.2597)(4.6779,4.1798)(4.7875,4.1134)
(4.9046,4.0615)(5.0275,4.0251)(5.154,4.0046)(5.2821,4.0005)(5.4096,4.0128)(5.5345,4.0413)(5.6548,4.0856)
(5.7684,4.1449)(5.8735,4.2182)(5.9683,4.3043)(6.0514,4.4019)(6.1213,4.5093)(6.1769,4.6247)(6.2173,4.7463)
(6.2418,4.8721)(6.25,5)
\put(3,6.75){\makebox(0,0){\small $S_1$}}
\put(7,6.75){\makebox(0,0){\small $S_2$}}
\put(3,5){\makebox(0,0){\small $C_1$}}
\put(7,5){\makebox(0,0){\small $C_2$}}
\put(4.75,4.75){\makebox(0,0){\small $L$}}
\put(4.2989,5.8069){\makebox(0,0){\small $v$}}
\put(5.1587,5.5026){\makebox(0,0){\small $u$}}
\put(4.537,5.6878){\makebox(0,0){\tiny$\bullet$}}
\put(4.9206,5.4894){\makebox(0,0){\tiny$\bullet$}}
\path(9.246,5.9921)(9.246,6.9709)
\path(9.1794,6.8554)(9.246,6.9709)(9.3127,6.8554)
\path(9.3127,6.1075)(9.246,5.9921)(9.1794,6.1075)
\path(9.25,4)(9.25,3)
\path(9.3167,3.1155)(9.25,3)(9.1833,3.1155)
\path(9.1833,3.8845)(9.25,4)(9.3167,3.8845)
\path(9.25,6)(9.25,4)
\path(9.3167,4.1155)(9.25,4)(9.1833,4.1155)
\path(9.1833,5.8845)(9.25,6)(9.3167,5.8845)
\put(9.5,6.5){\makebox(0,0){\small $s$}}
\put(9.5,3.5){\makebox(0,0){\small $s$}}
\put(9.5,5){\makebox(0,0){\small $2r$}}
\end{picture}
\caption{The Rolling Ball Method}
\label{RollingBall}
\end{figure}
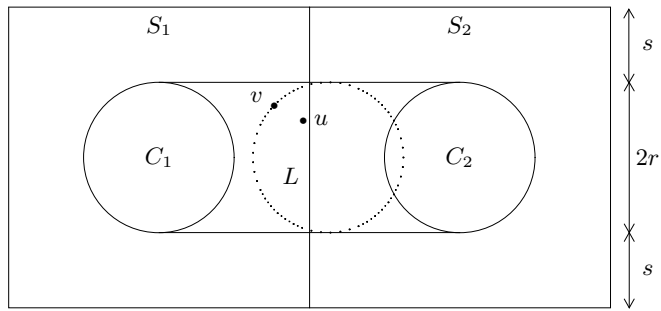

Moreover, such methods are based on comparison with independent bond percolation on the grid $\mathbb Z^2$. 
As, in Balister and Bollob\'as \cite{Balister10} in the case of $k$ nearest neighbor graph, we adapt the method of rolling ball 
(see figure \ref{RollingBall}) to the RNG. We procceed in two steps:
\begin{enumerate}
\item Controling the probability of some suitable configurations of points under Poisson point process. 
\item Consider the Bernoulli site percolation in such configurations.
\end{enumerate}

To prove that continuous percolation occurs, we shall compare the process to various bond percolation
models on $\mathbb{Z}^2$. In these models, the states of the edges will not be independent.
However they will satisfy the following definition:

\begin{definition} A bond percolation model is 1-independent if whenever $E_1$ and
$E_2$ are sets of edges at graph distance at least 1 from each another (i.e., if no edge
of $E_1$ is incident to an edge of $E_2$) then the state of the edges in $E_1$ is independent
of the state of the edges in $E_2$.
\end{definition}

We shall use the following result in Balister et al. \cite{Balister05}.

\begin{theorem} 
\label{TheoBalister}
If every edge in a 1-independent bond percolation model on $\mathbb{Z}^2$ is
open with probability at least $0.8639$, then almost surely there is an infinite open
component. Moreover, for any bounded region, there is almost surely a cycle of
open edges surrounding this region.
\end{theorem}

Let us first consider the case of percolation in the RNG. Write $u \sim v$ if $uv$ is an edge
of the underlying graph $\Rng(\varphi)$. For percolation we need to find an infinite path, i.e., a sequence $u_1,u_{2}\ldots$  with
$u_i \sim u_{i+1} $ for all $i$. Consider the rectangular region consisting of two adjacent
squares $S_1$, $S_2$ shown in figure \ref{RollingBall}. Both $S_1$ and $S_2$ have side length $2r+2s$, where
$r$ and $s$ are to be chosen later.  We define the basic
good event ${\cal E} _{S_1,S_2}$ to be the event that every vertex $u_1$ in the central disk $C_1$ of
$S_1$ is joined to at least one vertex $v$ in the central disk $C_2$ of $S_2$ by a $\Rng-$ path,
regardless of the state of the Poisson process outside of $S_1$ and $S_2$.

Now consider the following percolation model on $\mathbb{Z}^2$. Each vertex $(i, j)\in \mathbb{Z}^2$
corresponds to a square $[Ri,R(i + 1)] \times [Rj,R(j + 1)] \in \mathbb{R}^2$, where $R = 2r + 2s$,
and an edge is open between adjacent vertices (corresponding to squares $S_1$ and
$S_2$) if both the corresponding basic good events ${\cal E}_{S_1,S_2}$ and ${\cal E}_{S_{2},S_1}$ hold. Note
that this is indeed a 1-independent model on $\mathbb{Z}^2$ since the event ${\cal E}_{S_1,S_2}$ depends
only on the Poisson process within the region $S_1$ and $S_2$, and thus sets of edges at
distance at least one apart in $\mathbb{Z}^2$ depend on the Poisson process in disjoint regions
of $\mathbb {R}^2$. Any open path in $\mathbb{Z}^2$ corresponds to a sequence of basic good
events ${\cal E}_{S_1,S_2},{\cal E}_{S_{2},S_3}\ldots$ that occur, where $S_i$ is the square associated with a site in $\mathbb{Z}^2$. Every
vertex $u_1$ of the original Poisson process that lies in the central disk $C_1$ of $S_1$ now
has an infinite path leading away from it, since one can find points $u_i$ in the
central disk of $S_i$ and paths from $u_{i-1}$ to $u_i$ inductively for every $i > 1$. In
particular, each such $u_1$ lies in an infinite component. Moreover, such vertices
exist in $C_1$, so there is an infinite component.  One can choose
$r$ and $s$ so that the probability that the intersection of a basic good event is large and then we will apply the theorem \ref{TheoBalister}.

In order to bound the probability that this intersection of a basic good event fails, we shall use the
following rolling ball method. Let $C_1$, $C_2$, and $L$ be as in Figure \ref{RollingBall}. ($L$ is the
region between the two disks $C_1$ and $C_2$.) We need to define ${\bf E}_{S_1,S_2}$
 the event that for every point $v\in C_{1}\cup L$, there is a $u$ such that:\newline
 a) $ v \sim u$;\newline
 b) $d(u,v)\leq s$; and\newline
 c) $u \in D_{v}$, where $D_v$ is the disk of radius $r$ inside $C_{1}\cup L \cup C_{2}$ with $v$ on its
$C_1$-side boundary (the dotted disk in Figure \ref{RollingBall}).
Note in particular that (b) implies that the condition $u\sim v$ in (a) is independent
of the Poisson process outside of $S_1 \cup S_2$. This is because both $u$ and $v$ are at
distance at least $s$ from the exterior of $S_{1}\cup S_{2}$, so the event that $(u,v)$ is an edge of the RNG only
depends on the points within $S_{1}\cup S_2$. We denote $\bar{{\bf E}}_{S_1,S_2}$ the complementary of ${\bf E}_{S_1,S_2}$. 
The probability of $\bar{{\bf E}}_{S_1,S_2}$ is bounded by the expected number of points $u$ for which above conditions (a)-(c) fail. Thus, we have
\begin{equation}
\label{MajProb}
 \Pi(\bar{{\bf E}}_{S_1,S_2})\leq  2r(2r + 2s)p_{\Rng,r,s}
\end{equation}
where $p_{\Rng,r,s}$ is the probability that (a)-(c) fail for some fixed $v$. Notice that this
probability is independent of the location of $v$ in $C_{1}\cup L$.

\begin{lemma}
We can choose $r$ and $s$ such that  $\Pi(\bar{{\bf E}}_{S_1,S_2})$ is arbitrary small. 
\end{lemma}

{\bf Proof}:

Let $D(x,\alpha)$ be the disk of radius $\alpha$ and of center $x$.\\ 

Then,
$$p_{\Rng,\,r,\,s}\leq e^{ -|D_{v}\cap D(v,s)|} 
 + \int_{D_{v}\cap D(v,s)}p_{\,r,\,s}(u)\,du$$
where $p_{\,r,\,s}(u)=e^{-|D_{v}\cap D(v,\,d(u,v))|}
 (1- e^{-|D(v,\,d(u,v))\cap D(u,\,d(u,v))\backslash D_v|})$
 is the probability that $u$ is the closest point to $v$
inside $D_v$, but that  $(u,v)$ is not an edge of the RNG. To calculate this upper bound, note that
$$|D_{v}\cap D(v,s)|=-rs\sqrt{1-\frac{s^2}{4r^2}}+(2r^2-s^2)\arcsin\left(\frac{s}{2r}\right)+s^2\pi/2.$$
By choosing polar coordinates $(\alpha,\theta)$ of $u$, it comes: 

 $$\int_{D_{v}\cap D(v,s)}p_{\,r,\,s}(u)\,du=2\int_0^s\alpha e^{-|D_{v}\cap D(v,\alpha)|}\, \int_0^{\arccos(\alpha/(2r))} J(\alpha,\theta)\,d\theta\,d\alpha$$
where $$J(\alpha,\theta)=1-e^{-|D(v,\,\alpha)\cap D((\alpha,\theta),\,\alpha)\backslash D_v|}.$$

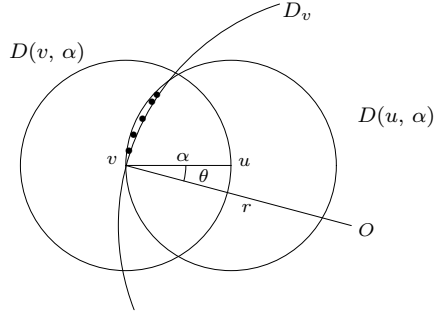
\begin{figure}
\centering
\unitlength 0.8cm
\begin{picture}(10,7)(0,2)%
\thinlines 
\put(3,5){\circle{3.5}}
\put(4.75,5){\circle{3.5}}
\path(4.75,5)(3,5)
\put(2.7778,5.0926){\makebox(0,0){\scriptsize $v$}}
\put(4.9735,5.0529){\makebox(0,0){\scriptsize $u$}}
\put(5.0132,4.2857){\makebox(0,0){\scriptsize $r$}}
\put(3.955,5.1455){\makebox(0,0){\scriptsize $\alpha$}}
\put(7,3.9286){\makebox(0,0){\scriptsize $O$}}
\path(3,5)(6.75,4)
\put(3,5){\arc{2}{0}{-6.0226}}
\put(4.2989,4.838){\makebox(0,0){\scriptsize $\theta$}}
\put(6.75,4){\arc{7.7478}{-3.5114}{-1.8842}}
\put(5.8598,7.5661){\makebox(0,0){\footnotesize $D_v$}}
\put(3.5185,6.1772){\makebox(0,0){\tiny$\bullet$}}
\put(3.1349,5.5026){\makebox(0,0){\tiny$\bullet$}}
\put(3.0556,5.2513){\makebox(0,0){\tiny$\bullet$}}
\put(3.4392,6.0582){\makebox(0,0){\tiny$\bullet$}}
\put(3.2804,5.7804){\makebox(0,0){\tiny$\bullet$}}
\put(1.7,6.8783){\makebox(0,0){\footnotesize $D(v,\,\alpha)$}}
\put(7.5,5.8069){\makebox(0,0){\footnotesize $D(u,\,\alpha)$}}
\end{picture}
\caption{The Lune $L(\alpha, \,r,\,\theta)= D(u,\,\alpha)\backslash D_v$ is the dotted area}
\label{essai}
\end{figure}

To calculate this last integration, we have :
$$\left\{\begin{array}{l}
\displaystyle\int_0^{\arccos\left(\frac{\alpha}{2r}\right)-\pi/3}J(\alpha,\theta)\,d\theta=\arccos\left(\frac{\alpha}{2r}\right)-\pi/3-\int_0^{\arccos\left(\frac{\alpha}{2r}\right)-\pi/3}e^{-|L(\alpha,\,r,\,\theta)|}\,d\theta\\
~\\
\displaystyle\int_{\arccos\left(\frac{\alpha}{2r}\right)-\pi/3}^{\arccos\left(\frac{\alpha}{2r}\right)}J(\alpha,\theta)\,d\theta=\pi/3- 2\displaystyle\frac{e^{-|L(\alpha,\,r,\,\arccos\left(\frac{\alpha}{2r}\right)-\pi/3)|}}{\alpha^2}\displaystyle\left[1-e^{-\alpha^2\pi/6}\right]
\end{array}\right.$$
with
$$
\left\{\begin{array}{l}
|L(\alpha,\,r,\,\theta)|=\alpha^2\theta+(\alpha^2-r^2)\arcsin\left(\displaystyle\frac{\alpha\sin\theta}{\sqrt{r^{2}+\alpha^{2} -2\alpha r\cos\theta}}\right)+\alpha r\sin\theta\\
~\\
|L(\alpha,r,\arccos\left(\frac{\alpha}{2r}\right)-\pi/3)|=\frac{\alpha^2}2(\frac{\pi}3-\frac{\sqrt{3}}2)-r^2\arcsin\left(\frac{\alpha}{2r}\right)+\frac{r\alpha}2\sqrt{1-\frac{\alpha^2}{4r^2}}.
\end{array}\right.
$$
Thus,
$$\begin{array}{ll}
\displaystyle\int_{D_{v}\cap D(v,s)}p_{\,r,\,s}(u)\,du&=2\displaystyle\int_0^{s}\alpha e^{-|D_{v}\cap D(v,\alpha)|}\times\left\{\arccos\left(\frac{\alpha}{2r}\right)-\right.\\
&\hspace*{-2.5cm}\left.\displaystyle\int_0^{\arccos\left(\frac{\alpha}{2r}\right)-\pi/3}e^{-|L(\alpha,\,r,\,\theta)|}\,d\theta-\displaystyle\frac{2e^{-|L(\alpha,\,r,\,\arccos\left(\frac{\alpha}{2r}\right)-\pi/3)|}}{\alpha^2}\displaystyle\left[1-e^{-\alpha^2\pi/6}\right]\right\}\,d\alpha.\end{array}$$
Take
$$\begin{array}{ll}
\displaystyle\int_{D_{v}\cap D(v,s)}p_{\,r,\,s}(u)\,du&=2\displaystyle\int_0^{s}\alpha \arccos\left(\frac{\alpha}{2r}\right) e^{-|D_{v}\cap D(v,\alpha)|}d\alpha\\
&-2\displaystyle\int_0^{s}\alpha e^{-|D_{v}\cap D(v,\alpha)|}\displaystyle\int_0^{\arccos\left(\frac{\alpha}{2r}\right)-\pi/3}e^{-|L(\alpha,r,\theta)|}\,d\theta\,d\alpha\\
&-4\displaystyle\int_0^{s}\frac1{\alpha}e^{-|D_{v}\cap D(v,\alpha)|-|L(\alpha,r,\arccos\left(\frac{\alpha}{2r}\right)-\pi/3)|}\displaystyle\left[1-e^{-\alpha^2\pi/6}\right]d\alpha
\end{array}$$
we conclude that 
$$\begin{array}{ll}
p_{\Rng,\,r,\,s}&
\leq e^{ -|D_{v}\cap D(v,s)|} 
 + \displaystyle\int_{D_{v}\cap D(v,s)}p_{\,r,\,s}(u)\,du\\
&= 1-2\displaystyle\int_0^{s}\alpha e^{-|D_{v}\cap D(v,\alpha)|}\displaystyle\int_0^{\arccos\left(\frac{\alpha}{2r}\right)-\pi/3}e^{-|L(\alpha,r,\theta)|}\,d\theta\,d\alpha\\
&-4\displaystyle\int_0^{s}\frac1{\alpha}e^{-|D_{v}\cap D(v,\alpha)|-|L(\alpha,r,\arccos\left(\frac{\alpha}{2r}\right)-\pi/3)|}\displaystyle\left[1-e^{-\alpha^2\pi/6}\right]\,d\alpha\\
\end{array}$$
which can be bounded by

\begin{eqnarray*}
\hspace*{-2cm}p_{\Rng,\,r,\,s}\leq & 1-2\displaystyle\int_0^{s}\alpha\, e^{-|D_{v}\cap D(v,\alpha)|-|L(\alpha,r,\arccos\left(\frac{\alpha}{2r}\right)-\pi/3)|}\\
&\times\left[\arccos\left(\frac{\alpha}{2r}\right)-\pi/3 +\frac{2(1-e^{-\alpha^2\pi/6})}{\alpha^2}\right]\,\,d\alpha.
\end{eqnarray*}
 
\begin{flushright}
$ \Box$
\end{flushright}

For instance, the bound involved in inequality \ref{MajProb} gives around $10^{-40}$ with $r=s=8000$.\\

For our purpose, we also need to control the probability for the Poisson point process of having at least one point in $C_1$ and less than $m$ points in $C_{2}\cup C_{1}\cup L$.

We denote $F_{C_1}=\{  \Phi(C_{1})\geq 1\}$ and $A_{m}=\{ \Phi(C_{2}\cup C_{1}\cup L)\leq m\}$.
Notice that as ${\bf E}_{S_1,S_2}\cap F_{C_1}\cap A_{m}\subset {\cal E}_{S_1,S_2}$
then $${\bf E}_{S_1,S_2}\cap {\bf E}_{S_2,S_1} \cap F_{C_1}\cap F_{C_2}\cap A_{m}\subset {\cal E}_{S_1,S_2}\cap {\cal E}_{S_2,S_1}.$$
We obtain that 
$$\Pi({\cal E}_{S_1,S_2}\cap{\cal E}_{S_{2},S_1})\geq 1-\left[\Pi(\bar{{\bf E}}_{S_1,S_2})+\Pi(\bar{{\bf E}}_{S_2,S_1})+\Pi(\bar{F}_{C_1})+ \Pi(\bar{F}_{C_2})+ \Pi(\bar{A}_{m})\right] $$
where $\Pi(\bar{F}_{C_1})=\Pi(\bar{F}_{C_2})=e^{-\pi r^{2}}$ and $$\Pi(\bar{A}_{m})=\displaystyle\sum_{k>m}\frac{ (2r(2r+2s)+\pi r^2)^{k}}{k!}e^{-2r(2r+2s)-\pi r^2}.$$
Choosing $r=s$, we have the following bound  $$\Pi(\bar{A}_{m})\leq \frac{((8+\pi)r^{2})^{m+1}}{(m+1)!}.$$ This bound  becomes negligible whenever $r\gg 1$ and $m> e(8+\pi)r^{2}$ using Stirling formula.\\

Set $\epsilon=0.1361$.
Similarly as in H{\"a}ggstr{\"o}m~\cite{Haggstrom00} but adapted in the 1-independent case, let $B_{r,s}$ be the event that
all the sites are open in $C_{2}\cup C_{1}\cup L$ with probability 
$p=1-\frac{\epsilon}{2m}$ and also define 
$C_{r,s}= {\cal E}_{S_1,S_2}\cap{\cal E}_{S_{2},S_1}\cap B_{r,s}$. Then, we have:
$$\begin{array}{ll}
\mathbb{P}\left(C_{r,s}\right)&= \displaystyle\int_{{\cal E}_{S_1,S_2}\cap {\cal E}_{S_{2},S_1}}\Pi\left(d\varphi\right)\mu_\varphi^{p}\left( B_{r,s}\right)\\
&\geq \displaystyle\int_{{\bf E}_{S_1,S_2}\cap {\bf E}_{S_2,S_1} \cap F_{C_1}\cap F_{C_2}\cap A_{m} }\Pi\left(d\varphi\right)\mu_\varphi^{p}\left( B_{r,s}\right)\\
&~\\
&\geq (1-\epsilon/2)p^{m}> 1-\epsilon=0.8639
\end{array}$$
because we can choose $r,s,m$ (as preceding) such that
 $$\Pi({\bf E}_{S_1,S_2}\cap {\bf E}_{S_2,S_1} \cap F_{C_1}\cap F_{C_2}\cap A_{m})\geq 1-\epsilon/2$$
We conclude with theorem \ref{TheoBalister} that $p_c^{\mathrm{site}}(\Rng,\Pi)\leq1-\frac{\epsilon}{2m}<1$.

\section{Concluding remarks}

This kind of proof also apply in the case of bond or site percolation of the $k$ nearest neighbor graph: it is direct consequence of \cite{Balister10}. It is sufficient to use what they called $p_{U}$, $p_{B}$, $p_{I}$ or $p_{O}$ for several possible definitions of percolation.

As suggested in \cite{Balister10} with a high confidence, $k=3$ is the critical out-degrees for percolation on the $k$ nearest neighbor graph. We notice that in the RNG the number of neighbors is bounded by $6$. So a point have neighbors in several directions. It may be interesting to study a family of Delaunay subgraphs defined on sequence of vacuity regions such that the number of neighbors of each point is the lowest as possible but keep good connectivity properties for percolation purposes. Otherwise a challenge would be to extend the method of the rolling ball when points are distributed under a Gibbs point process for instance a hard-core point process.

\bibliographystyle{plain}
\bibliography{bbd}

\end{document}